\documentclass[twocolumn,preprintnumbers,amsmath,amssymb]{revtex4}

\usepackage{graphicx}
\usepackage{dcolumn}
\usepackage{bm}

\begin{document}


\title{Generalized friendship paradox in complex networks:
\\ The case of scientific collaboration}

\author{Young-Ho Eom}
\affiliation{{\it Laboratoire de Physique Th\'eorique du CNRS,
IRSAMC, Universit\'e de Toulouse, UPS, F-31062 Toulouse, France}}
\author{Hang-Hyun Jo}
\affiliation{BECS, Aalto University School of Science, P.O. Box
12200, Espoo, Finland}


\begin{abstract}
The friendship paradox states that your friends have on average
more friends than you have. Does the paradox ``hold'' for other
individual characteristics like income or happiness? To address
this question, we generalize the friendship paradox for arbitrary
node characteristics in complex networks. By analyzing two
coauthorship networks of Physical Review journals and Google
Scholar profiles, we find that the generalized friendship paradox
(GFP) holds at the individual and network levels for various
characteristics, including the number of coauthors, the number of
citations, and the number of publications. The origin of the GFP
is shown to be rooted in positive correlations between degree and
characteristics. As a fruitful application of the GFP, we suggest
effective and efficient sampling methods for identifying high
characteristic nodes in large-scale networks. Our study on the GFP
can shed lights on understanding the interplay between network
structure and node characteristics in complex networks.
\end{abstract}

\maketitle

\section*{Introduction}

People live in social networks. Various behaviors of individuals
are significantly influenced by their positions in such networks,
whether they are offline or
online~\cite{Watts1998,Castello2009,Lazer2009}. Through the
interaction and communication among individuals, information,
behaviors, and diseases
spread~\cite{Vespignani2011,Bakshy2012,Centola2010,Christakis2007,Pastor-Satorras2001,Weng2013,Marvel2013}.
Thus understanding the structure of social networks could enable
us to understand, predict, and even control social collective
behaviors taking place on or via those networks. Social networks
have been known to be heterogeneous, characterized by broad
distributions of the number of neighbors or
degree~\cite{Barabasi1998}, assortative mixing~\cite{Newman2002},
and community structure~\cite{Fortunato2010} to name a few.

One of interesting phenomena due to the structural heterogeneity
in social networks is the friendship paradox~\cite{Feld1991}. The
friendship paradox (FP) can be formulated at individual and
network levels, respectively. At the individual level, the paradox
holds for a node if the node has smaller degree than the average
degree of its neighbors. It has been shown that the paradox holds
for most of nodes in both offline and online social
networks~\cite{Feld1991,Ugander2011,Hodas2013}. However, most
people believe that they have more friends than their friends
have~\cite{Zuckerman2001}. The paradox holds for a network if the
average degree of nodes in the network is smaller than the average
degree of their neighbors~\cite{Feld1991}. The paradox can be
understood as a sampling bias in which individuals having more
friends are more likely to be observed by their friends. This bias
has important implications for the dynamical processes on social
networks, especially when it is crucial for the process to
identify individuals having many neighbors, or high degree nodes.
For example, let us consider the spreading process on networks. It
turns out that sampling neighbors of random individuals is more
effective and efficient than sampling random individuals for the
early detection of epidemic spreading in large-scale social
networks~\cite{Christakis2010,Garcia-Herranz2012}, and for
developing efficient immunization strategies in computer
networks~\cite{Cohen2003}. Recently, the information overwhelming
or spam in social networking services like
Twitter~\cite{Hodas2013} has been also explained in terms of the
friendship paradox.

The friendship paradox has been considered only as the topological
structure of social networks, mainly by focusing on the number of
neighbors, among many other node characteristics. Each individual
could be described by his/her cultural background, gender, age,
job, personal interests, and genetic
information~\cite{Park2007,Fowler2008}. This is also the case for
other kinds of networks: Web pages have their own fitness in World
Wide Web~\cite{Kong2008}, and scientific papers have intrinsic
attractiveness in a citation network~\cite{Eom2011}. These
characteristics play significant roles in dynamical processes on
complex networks~\cite{Park2007,Fowler2008, Kong2008, Eom2011,
Aral2009}. Hence, one can ask the question: Can the friendship paradox be applied to node characteristics other than degree?

To address this question, we generalize the friendship paradox for
arbitrary node characteristics including degree. Similarly to the
FP, our generalized friendship paradox (GFP) can be formulated at
individual and network levels. The GFP holds for a node if the
node has lower characteristic than the average characteristic of
its neighbors. The GFP holds for a network if the average
characteristic of nodes in the network is smaller than the average
characteristic of their neighbors. When the degree is considered
as the node characteristic, the GFP reduces to the FP. In this
paper, by analyzing two coauthorship networks of physicists and of
network scientists, we show that your coauthors have more
coauthors, more citations, and more publications than you have.
This indicates that the friendship paradox holds not only for
degree but also for other node characteristics. We also provide a
simple analysis to show that the origin of the GFP is rooted in
the positive correlation between degree and node characteristics.
As applications of the GFP, two sampling methods are suggested for
sampling nodes with high characteristics. We show that these
methods are simple yet effective and efficient in large-scale
social networks.

\section*{Results}

\subsection*{Generalized friendship paradox in complex networks}

We consider two coauthorship networks constructed from the
bibliographic information of Physical Review (PR) journals and
Google Scholar (GS) profile dataset of network scientists (See
Method Section). Each node of a network denotes an author of
papers and a link is established between two authors if they wrote
a paper together. The number of nodes, denoted by $N$, is 242592
for the PR network and 29968 for the GS network. For the node
characteristics in the PR network, we consider the number of
coauthors, the number of citations, the number of publications,
and the average number of citations per publication. As for the GS
network, the number of coauthors and the number of citations are
considered. The characteristic of node $i$ will be denoted by
$x_i$, and for the degree we denote it by $k_i$.

The generalized friendship paradox (GFP) can be studied at two
different levels: (i) Individual level and (ii) network level.

\begin{figure*}[ht]
\begin{center}
\includegraphics[width=130mm,angle=0]{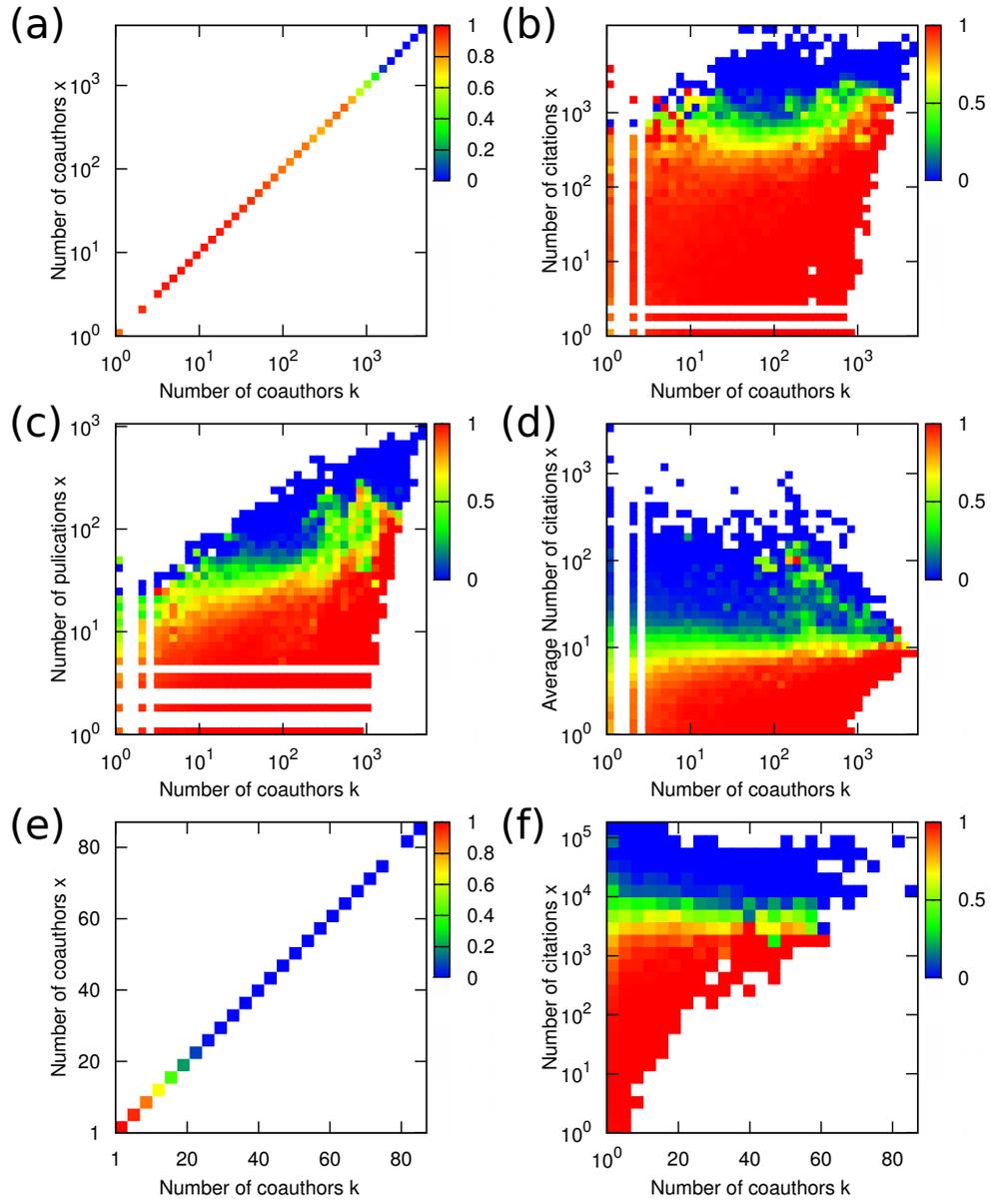}
\caption{The paradox holding probability $h(k,x)$ as a function of degree $k$
and node characteristic $x$. For the Physical Review (PR) coauthorship network,
we use (a) the number of coauthors, i.e., $x=k$, (b) the number of citations,
(c) the number of publication, and (d) the average number of citations
per publication, while for the Google Scholar (GS) coauthorship network,
we use (e) the number of coauthors, i.e., $x=k$, and (f) the number of citations.}
\label{fig:RKX}
\end{center}
\end{figure*}

(i) \emph{Individual level.} The GFP holds for a node $i$ if the
following condition is satisfied:
\begin{equation}
x_i < \frac{\sum_{j \in {\Lambda}_i}x_j}{k_i},
\label{eq:NodeParadox}
\end{equation}
where ${\Lambda}_i$ denotes the set of neighbors of node $i$. Note
that setting $x_i=k_i$ reduces the GFP to the FP. We define the
paradox holding probability $h(k,x)$ that a node with degree $k$
and characteristic $x$ satisfies the condition in
Eq.~(\ref{eq:NodeParadox}). Figure~\ref{fig:RKX} shows the
empirical results of $h(k,x)$ for PR and GS networks. It is found
that for fixed degree $k$, $h(k,x)$ decreases with increasing $x$
for any characteristic $x$ other than $k$
(Fig.~\ref{fig:RKX}~(b--d,f)). The same decreasing tendency has
been observed for $x=k$ (Fig.~\ref{fig:RKX}~(a,e)). In
Eq.~(\ref{eq:NodeParadox}), the larger value of $x_i$ is expected
to lower the probability $h(k,x)$ if the characteristics of node
$i$'s neighbors remain the same. As a limiting case, the node with
minimum value of $x$, i.e., $x_{\rm min}$, is most likely to have
friends with higher values of $x$, leading to $h(k,x_{\rm
min})=1$. On the other hand, for the node with maximum value of
$x$, we get $h(k,x_{\rm max})=0$.

Next, the dependence of $h(k,x)$ on the degree $k$ can be
classified as either increasing or being constant. Here the case
of $x$ denoting the degree is disregarded for both networks. The
increasing behavior is observed mainly for the number of citations
and the number of publications in the PR network in
Fig.~\ref{fig:RKX} (b,c), while the constant behavior is observed
for the average number of citations per publication in the PR
network and for the number of citations in the GS network, shown
in Fig.~\ref{fig:RKX} (d,f), respectively. In order to understand
such difference, we calculate the Pearson correlation coefficient
between $k$ and $x$ as
\begin{equation}
  \rho_{kx} =\frac{1}{N}\sum_{i=1}^N \frac{(k_i - \langle k \rangle)(x_i - \langle x \rangle)}{\sigma_{k}\sigma_{x}},
\end{equation}
where $\langle x\rangle$ and $\sigma_x$ denote the average and
standard deviation of $x$. We also obtain the characteristic
assortativity for each characteristic $x$, adopted
from~\cite{Newman2002}:
\begin{equation}
  r_{xx} = \frac{L\sum_l x_lx'_l-[\sum_l\frac{1}{2}(x_l+x'_l)]^2} {L\sum_l\frac{1}{2}({x_l}^2+{x'_l}^2) -[\sum_l\frac{1}{2}(x_l+x'_l)]^2},
\end{equation}
where $x_l$ and $x'_l$ denote characteristics of nodes of the
$l$th link, with $l=1,\cdots,L$ and $L$ is the total number of
links in the network. The value of $r_{xx}$ ranges from $-1$ to
$1$, and it increases according to the tendency of high
characteristic nodes to be connected to other high characteristic
nodes. The values of these quantities are summarized in Table I.
From now on, we denote the degree assortativity as $r_{kk}$.

The $k$-dependent behavior of $h(k,x)$ can be understood mainly as
the combined effect of $r_{kk}$ and $\rho_{kx}$. Since
$r_{kk}\approx 0.47$ in the PR network, for a node $i$ with fixed
$x_i$, the larger $k_i$ implies the larger $k_j$ of its friend
$j$. This may lead to the higher $x_j$, e.g., due to
$\rho_{kx}\approx 0.79$ for the number of publications, leading to
the increasing behavior of $h(k,x)$. However, for the average
number of citations per publication showing $\rho_{kx}\approx
0.07$, the larger $k_j$ does not imply the higher $x_j$, which
leads to the constant behavior of $h(k,x)$. For the number of
citations in the GS network, the almost neutral degree correlation
by $r_{kk}\approx -0.02$ inhibits any correlated behavior between
characteristics, thus we again observe the constant behavior of
$h(k,x)$. We note that the neutral degree correlation in the GS
network is unlike many other coauthorship networks, mainly due to
incomplete information available from GS profiles, and due to the
snowball sampling method we employed~\cite{Lee2006}.

Now we define the average paradox holding probability as $H=\sum_k
\int dx h(k,x)P(k,x)$, where $P(k,x)$ denotes the probability
distribution function of node with degree $k$ and characteristic
$x$. As shown in Table I, the value of $H$ is larger than $0.7$
for every considered characteristic, implying that the GFP holds
at the individual level to a large extent.

(ii) \emph{Network level.} In order to investigate the GFP at the
network level, we define the average characteristic of neighbors
$\langle x\rangle_{nn}$ for comparing it to the average
characteristic $\langle x\rangle$:
\begin{equation}
\langle x \rangle_{nn} = \frac{\sum_{i=1}^N k_i x_i}{\sum_{i=1}^N k_i}.
\end{equation}
Here a node $i$ with degree $k_i$ has been considered as a
neighbor $k_i$ times. The GFP holds at the network level if the
following condition is satisfied:
\begin{equation}
\langle x \rangle < \langle x \rangle_{nn}.
\end{equation}
Note that setting $x_i=k_i$ reduces the GFP to the FP. As shown in
Table I, the GFP holds for all characteristics considered. In
other words, your coauthors have on average more coauthors, more
citations, and more publications than you have.

In summary, our results indicate that the generalized friendship
paradox holds at both individual and network levels for many node
characteristics of networks.

\subsection*{Origin of the GFP}

The prevalence of the GFP for most nodes in networks regardless of
node characteristics implies that there might be a universal
origin of the GFP. For the original friendship paradox, the
existence of hub nodes and the variance of degree have been
suggested for the origin of the paradox~\cite{Feld1991}. In order
to investigate the origin of the GFP at the network level, we
define a function $F=\langle x\rangle_{nn}-\langle x\rangle$, and
straightforwardly obtain the following equation:
\begin{equation}
  F=\langle x\rangle_{nn}-\langle x\rangle=\frac{\rho_{kx}\sigma_k\sigma_x}{\langle k\rangle}.
\end{equation}
One can say that the GFP holds if $F>0$. Since standard deviations
$\sigma_k$ and $\sigma_x$ are positive in any non-trivial cases,
the GFP holds if $\rho_{kx}>0$. Thus the degree-characteristic
correlation $\rho_{kx}$ is the key element for the generalized
friendship paradox. Note that in case when $x_i=k_i$, i.e.,
$\rho_{kk}=1$, the FP holds in any non-trivial cases.

The origin of the GFP can help us to better understand the
dynamical processes on networks when the characteristic $x$ is
considered to be a node activity such as communication frequency
or traffic. The positive correlation between degree and node
activity has been observed in mobile phone call
patterns~\cite{Onnela2007} and the air-transportation
network~\cite{Barrat2004}, enabling the application of the GFP to
those phenomena. In case of protein interaction networks, the
degrees of proteins are positively correlated with their
lethality~\cite{Jeong2001,Zotenko2008}, while they are negatively
correlated with their rates of evolution~\cite{Fraser2002}. The
negative degree-characteristic correlations, i.e., $\rho_{kx}<0$,
can lead to the opposite behavior of the GFP, which can be called
anti-GFP.

\subsection*{Sampling high characteristic nodes using GFP in complex networks}

Identifying important or central nodes in a network is crucial for
understanding the structure of complex networks and dynamical
processes on those networks. The recent advance of
information-communication technology (ICT) has opened up access to
the data on large-scale social networks. However, complete mapping
of social networks is not feasible, partially due to privacy
issues. Thus it is still important to devise proper sampling
methods that exploit local network structure. In this sense, the
original friendship paradox has been used to sample high degree
nodes in empirical networks. It was found that the set of neighbors
of randomly chosen nodes can have the predictive power of epidemic
spreading on both offline social networks~\cite{Christakis2010}
and online social networks~\cite{Garcia-Herranz2012}.

We suggest two simple sampling methods using the GFP to identify
high characteristic nodes in a network: (i) Friend sampling and
(ii) biased sampling. These methods are then compared to the
random sampling method to test whether our methods are more
efficient to sample high characteristic nodes. We first choose
random nodes to make a control group. For each node in the control
group, one of its neighbors is randomly chosen. These chosen nodes
compose a friend group. Finally, for each node in the control
group, we choose its neighbor having the highest characteristic to
make a biased group. For the biased sampling, we have assumed that
each node has the full information about characteristics of its
neighbors.

\begin{figure*}[hbt!]
\begin{center}
\includegraphics[width=120mm,angle=-90]{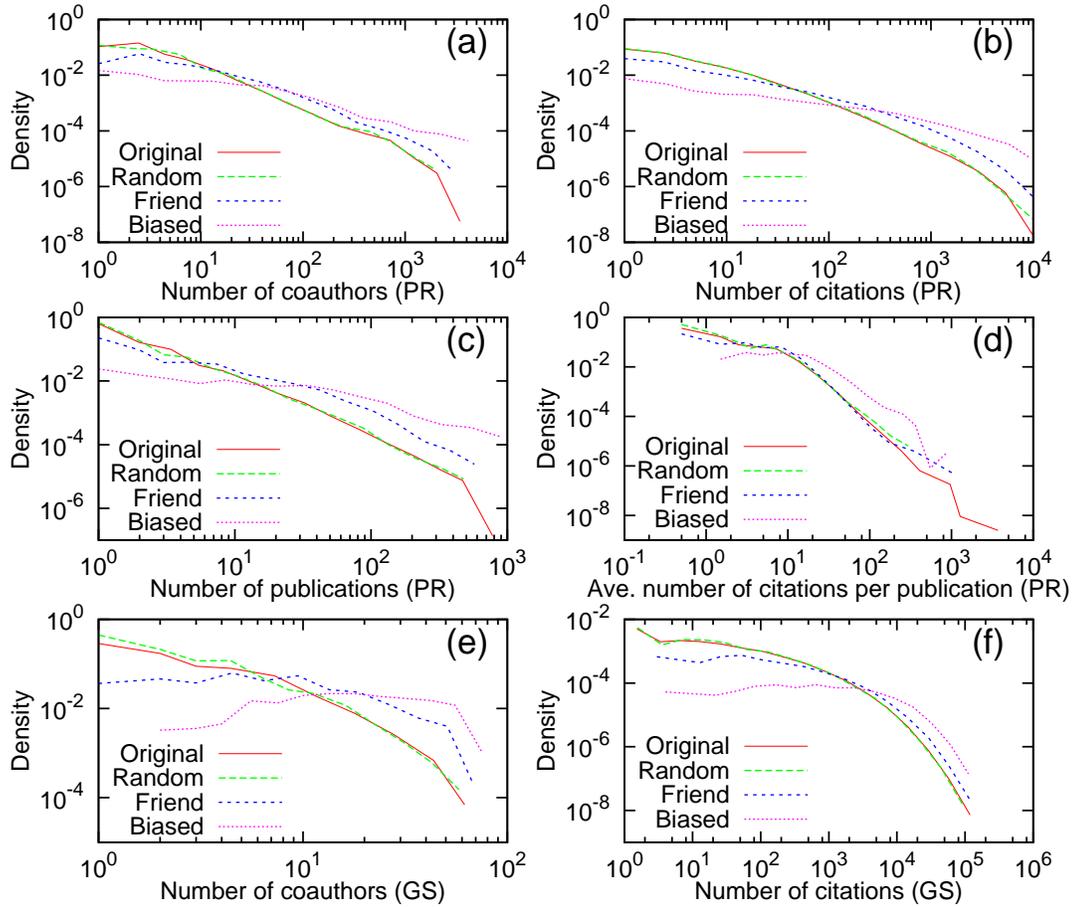}
\caption{Characteristic distributions for control group, friend
group, and biased group, for each of which 5000 nodes are sampled.
The original full distributions are also plotted for comparison.
We use (a) the number of coauthors (PR), (b) the number of
citations (PR), (c) the number of publications (PR), (d) the
average number of citations per publication (PR), (e) the number
of coauthors (GS), and (f) the number of citations
(GS).}\label{fig:SamplePR}
\end{center}
\end{figure*}

Figure~\ref{fig:SamplePR} shows the characteristic distributions
of sampled nodes from PR and GS networks by different sampling
methods. Heavier tails of distributions imply better sampling for
identifying high characteristic nodes. The performance of biased
sampling is the best in all cases because this sampling utilizes
more information about neighbors than the friend sampling. The
friend sampling shows better performance than the random sampling
(control group) for most characteristics as it is expected by
large values of $\rho_{kx}$. One exceptional case is for the
average number of citations per publication in the PR network,
shown in Fig.~\ref{fig:SamplePR} (d). Here the friend sampling
does not better than the random sampling due to the very small
degree-characteristic correlation, $\rho_{kx}\approx 0.07$, while
the result by biased sampling is still better than those by other
sampling methods.

\begin{figure*}[hbt!]
\begin{center}
\includegraphics[width=110mm,angle=-90]{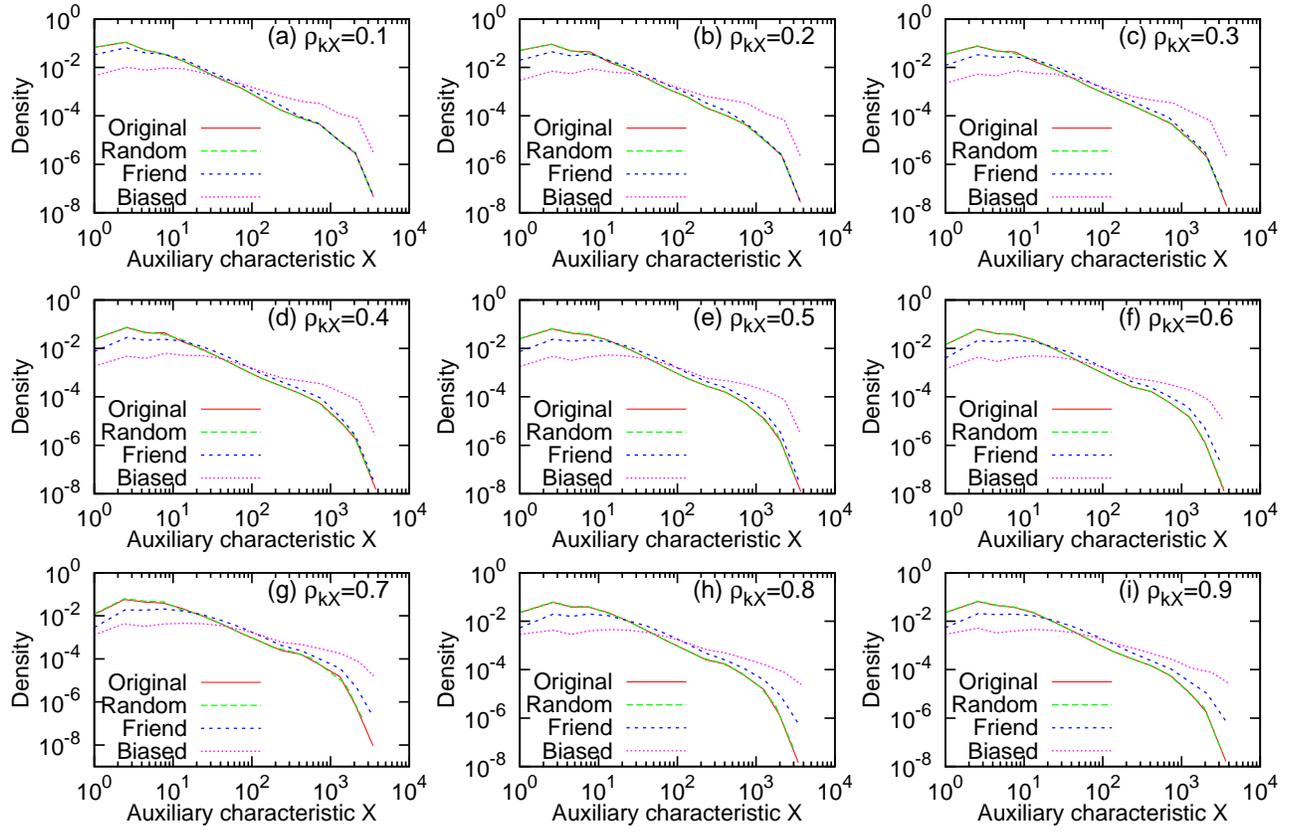}
  \caption{Performance comparison of control group, friend group,
  and biased group for auxiliary characteristics $X$ with various values of
  correlation with degree: $\rho_{kX}=0.1,\cdots,0.9$. For each value of $\rho_{kX}$,
  1000 random configurations are generated, for each of which 5000 nodes
  are sampled as in Fig.~\ref{fig:SamplePR}.}\label{fig:SampleCorrelation}
\end{center}
\end{figure*}

Next, in order to investigate the effect of degree-characteristic
correlation on the performance of sampling methods, we consider an
auxiliary characteristic $X$ based on the method of Cholesky
decomposition~\cite{Press1992}. To each node $i$ with degree $k_i$
in the PR network, we assign a characteristic $X_i$ given by
\begin{equation}
X_i = \rho k_i + \sqrt{1-\rho^2}y_i,
\end{equation}
where $y_i$ denotes the $i$th element of the shuffled set of
$\{k_i\}$. Since $\rho=\rho_{kX}$ (See Method Section), the
correlation can be easily controlled by $\rho$. Then we apply the
same sampling methods to identify nodes with high $X$, and compare
their performances for different values of $\rho_{kX}$.
Figure~\ref{fig:SampleCorrelation} shows that the biased sampling
performs significantly better than any other sampling methods,
independent of $\rho_{kX}$. The friend sampling performs better
than the random sampling, while the difference in performance
increases with the value of $\rho_{kX}$.

The sampling results suggest that the biased sampling can be very
efficient and effective to detect a group of high characteristic
nodes when the information about characteristics of neighbors is
available. Otherwise the friend sampling still performs better
than the random sampling.

\section*{Discussion}

Node characteristics have profound influence on the evolution of
networks~\cite{Kong2008,Eom2011} and dynamical processes on such
networks like
spreading~\cite{Aral2009,Christakis2010,Garcia-Herranz2012}. By
taking into account various node characteristics, we have
generalized the friendship paradox in complex networks. The
generalized friendship paradox (GFP) states that your friends have
on average higher characteristics than you have. By analyzing two
coauthorship networks of Physical Review (PR) journals and of
Google Scholar (GS) profiles, we have found that the GFP holds at
both individual and network levels for various node
characteristics, such as the number of coauthors, the number of
citations, the number of publications, and the average number of
citations per publication. It is also shown that the origin of the
GFP at the network level is rooted in the positive correlation
between degree and characteristic. Thus the GFP is expected to
hold for any characteristic showing the positive correlation with
degree. Here the characteristic can be also purely topological
like various node centralities as they show significant positive
correlations with degree, such as PageRank~\cite{Fortunato2008}.

Despite the access to the data on large-scale social networks,
complete mapping of social networks is not feasible. Thus it is
still important to devise effective and efficient sampling methods
that exploit local network structure. We have suggested two simple
sampling methods for identifying high characteristic nodes using
the GFP. It is empirically found that a control group of randomly
chosen nodes has the smaller number of high characteristic nodes
than a friend group that consists of random neighbors of nodes in the control
group. Moreover, provided that nodes have full information about
characteristics of their neighbors, a biased group of the highest
characteristic neighbors of nodes in the control group has the
largest number of high characteristic nodes than other groups.
This turns out to be the case even when the degree-characteristic
correlation is negligible.

Our sampling methods propose an explanation about how our
perception can be affected by our friends. People's perception of
the world and themselves depends on the status of their friends,
colleagues, and peers~\cite{Zuckerman2001}. When we compare our
characteristics like popularity, income, reputation, or happiness
to those of our friends, our perception of ourselves might be
distorted as expected by the GFP. Comparing to the average friend,
i.e., the friend sampling, is biased due to the positive
degree-characteristic correlation. Furthermore, comparing to the
``better'' friend, i.e., the biased sampling, is much more biased
towards the ``worse'' perception of ourselves. This might be the reason why
active online social networking service users are not
happy~\cite{Kross2013}, in which it is much easier to compare to
other people in online social media.

Another interesting application of the GFP can be found in
multiplex networks~\cite{Kivela2013,Jo2006}. If degrees of one
layer are positively correlated with those of other layers, our
sampling methods can be used to identify high degree nodes in
other layers. Indeed, the degrees of each node are positively
correlated across layers in a player network of an online
game~\cite{Szell2010} and in a multiplex transportation
network~\cite{Parshani2010}.

Nodes are not only embedded in the topological structure, but they
also have many other characteristics relevant to the structure and
evolution of complex networks. However, the role of these
non-topological characteristics is far from being fully
understood. Our work on the generalized friendship paradox will
help us consider the interplay between network structure and node
characteristics for deeper understanding of complex networks.

\section*{Methods}

\subsection*{Data description}

We describe how the data for coauthorship networks have been
collected and prepared. For the Physical Review (PR) network, the
bibliographic data containing all papers published in Physical
Review journals from 1893 to 2009 was downloaded from American
Physical Society. The number of papers is 463348, and each paper
has the title, the list of authors, the date of publication, and
citation information. By using author identification algorithm
proposed by~\cite{Radicchi2009}, we identified each author by
his/her last name and initials of first and middle names if
available. The number of identified authors is 242592. Combined
with the numbers of citations and the list of authors of papers,
we obtained for each author the number of coauthors, the number of
citations, the number of publications, and the average number of
citations per publication.

Google Scholar (GS) service (scholar.google.com) provides profiles
of academic authors. Each profile of the author contains
information of the total number of citations and coauthor list of
the author. Using snowball sampling~\cite{Lee2006} starting from
``Albert-L\'aszl\'o Barab\'asi'' (one of the leading network
scientists), the coauthor relations and their citation information
are collected. The number of authors in the dataset is 29968. Here
we note that not all scientists have profile in the GS and not all
coauthor relations are accessible.

\subsection*{Generating random node characteristics of arbitrary correlation with degree}

Consider two independent random variables $Y=(y_1,y_2,\ldots,
y_N)$ and $Z=(z_1,z_2,\ldots, z_N)$ with the same standard
deviation, i.e., $\sigma_{Y}=\sigma_{Z}$. We generate a random
sequence $X=(x_1,x_2,\ldots, x_N)$ from the following equation:
\begin{equation}
X = \rho Y + \sqrt{1-\rho^2}Z.
\end{equation}
The correlation $\rho_{XY}$ between $X$ and $Y$ is given by
\begin{equation}
\rho_{XY} = \frac{E(XY)-E(X)E(Y)}{\sigma_X \sigma_Y},
\end{equation}
where $E(X)$ denotes the expectation of $X$. Using the independence of $Y$ and $Z$, i.e.,
$E(YZ)=E(Y)E(Z)$, we get
\begin{equation}
\rho_{XY} = \rho \frac{\sigma_Y}{\sigma_X}.
\end{equation}
Then, from $\sigma_X^2 = E(X^2)-E(X)^2$, we obtain $\sigma_X=\sigma_Y$, leading to
\begin{equation}
\rho_{XY} = \rho.
\end{equation}





\section*{Acknowledgements}

The authors thank Daniel Kim and Hawoong Jeong for providing
Google Scholar data and American Physical Society for providing
Physical Review bibliographic data. Y.-H.E. acknowledges support
from the EC FET Open project ``New tools and algorithms for
directed network analysis'' (NADINE number 288956). H.-H.J.
acknowledges financial support by the Aalto University
postdoctoral programme.

\section*{Author Contribution}

Y.-H.E. and H.-H.J. designed research, wrote, reviewed, and
approved the manuscript. Y.-H.E. performed data collection and
analysis.

\section*{Additional Information}
Competing financial interests: The authors declare no competing
financial interests.
{}


\begin{table*}[!ht]
  \caption{Empirical results for the generalized friendship paradox in
   two coauthorship networks from Physical Review (PR) journals and from Google Scholar (GS) profiles.
   For each node characteristic $x$, we measure the Pearson correlation coefficient with degree $\rho_{kx}$,
   the characteristic assortativity $r_{xx}$, the average paradox holding probability $H$,
   and average characteristics of nodes $\langle x\rangle$ and their neighbors $\langle x\rangle_{nn}$.}
\label{table:EmpiResult}
\begin{center}
\begin{tabular}{|c|c|c|c|c|c|c|}
  \hline
characteristic $x$ & $\rho_{kx}$ & $r_{xx}$ & $H$ & $\langle x \rangle$ & & $\langle x\rangle_{nn}$ \\
  \hline
The number of coauthors (PR) & 1.00 & 0.47 & 0.934 & 58.3  & $<$ & 771.7 \\
The number of citations (PR) & 0.69 & 0.21 & 0.921 & 110.1 & $<$ &1135.7\\
The number of publications (PR) & 0.79 & 0.25 & 0.912 & 10.2  & $<$ &102.1\\
The average number of citations per publication (PR) & 0.07 & 0.34 & 0.720 & 7.8 & $<$ &12.4 \\
  \hline
The number of coauthors (GS) & 1.00 & -0.02 & 0.863 & 6.9 & $<$ & 16.1 \\
The number of citations (GS) & 0.44 & 0.14 & 0.792 & 3089.8 & $<$ & 5401.0 \\
\hline
\end{tabular}
\end{center}
\end{table*}

\end{document}